\documentclass[prd,aps,twocolumn,superscriptaddress,floatfix]{revtex4}
\usepackage{graphicx}
\usepackage{xcolor}
\usepackage{amsfonts}
\usepackage{amssymb,amsmath,bm}
\usepackage{hyperref}
\usepackage[export]{adjustbox}
\hypersetup{colorlinks=true,citecolor=blue}
\usepackage{multirow}

\hypersetup{
    colorlinks=false,
    citecolor=blue,
    linkcolor=blue,
    linktoc=page
}

\newcommand{\aap}{{Astron.~Astrophys.}}
\newcommand{\apjl}{{Astrophys.~J.~Lett.}}
\newcommand{\apjs}{{Astrophys.~J.~Supp.}}

\newcommand{\Msun}{\ensuremath{M_{\odot}}}

\newcommand{\avg}[1]{\ensuremath{\left\langle \,#1\, \right\rangle}}

\newcommand{\be}{\begin{equation}}
\newcommand{\ee}{\end{equation}}

\newcommand{\bea}{\begin{eqnarray}}
\newcommand{\eea}{\end{eqnarray}}
\newcommand{\bdm}{\begin{displaymath}}
\newcommand{\edm}{\end{displaymath}}

\def\Mpc{\, h^{-1} \, {\rm Mpc}}

\def\Gpc{\, h^{-1} \, {\rm Gpc}}

\def\kMpc{\, h \, {\rm Mpc}^{-1}}

\def\dk{\frac{\mathrm{d}^3\,k}{(2\pi)^3}\,}

\def\fNLL{f_{\mathrm{NL}}^{\mathrm{loc}}}

\newcommand{\fig}[1]{Figure~\ref{#1}}

\def\ie{{\em i.e.}~}
\def\eg{{{\em e.g.}~}}

\begin{document}

\title{Primordial non-Gaussianities and zero bias tracers of the Large Scale Structure}

\author{Emanuele Castorina}
\author{Yu Feng}
\author{Uro\v s Seljak}
\affiliation{Berkeley Center for Cosmological Physics, University of California, Berkeley, CA 94720}
\affiliation{Lawrence Berkeley National Laboratory, 1 Cyclotron Road, Berkeley, CA 93720, USA}
\author{Francisco Villaescusa-Navarro}
\affiliation{Center for Computational Astrophysics, Flatiron Institute, 162 5th Avenue, 10010, New York, NY, USA}
\begin{abstract}
We develop a new method to constraint primordial non-Gaussianities of the local kind using unclustered tracers of the Large Scale Structure. We show that in the limit of low noise, zero bias tracers yield large improvement over standard methods, mostly due to vanishing sampling variance. We propose a simple technique to construct such a tracer, using environmental information obtained from the original sample, and validate our method with N-body simulations. 
Our results indicate that $\sigma_{\fNLL}\simeq1$ can be reached using only information on a single tracer of sufficiently high number density.
\end{abstract}
\maketitle
\section{Introduction}

Understanding the initial conditions of the Universe is major open problem in theoretical cosmology.
The statistical properties of the primordial curvature perturbations are a key ingredient of the success of the $\Lambda {\rm CDM}$ model to explain the Universe as we observe it today. In the simplest models of inflation\citep{Starobinsky,Guth,Linde}, slow-roll single field inflation, initial fluctuations are Guassian for all practical purposes\citep{Maldacena03,Creminelli04,dePutter2017}, but current observations still allow a large variety of models predicting large Primordial Non-Gaussianities (PNG). This would be for instance the case if cosmological perturbations are not generated by the inflationary clock driving inflation, but rather by other fields\citep{Linde97,Lyth02,Dvali04,Zaldarriaga03,Sasaki2006}. This class of models often goes under the name of multi-field inflation. PNG contributing mostly to squeezed configurations of the primordial curvature bispectrum are called of the local kind. In terms of the primordial gravitational potential $\Phi(\mathbf{x})$, they can be parametrized with a single number $\fNLL$, $\Phi(\mathbf{x}) = \phi_g(\mathbf{x})+\fNLL(\phi_g(\mathbf{x})^2-\avg{\phi_g^2})$, with $\phi_g$ a Gaussian random field.

A general prediction of multi-field models is $|\fNLL|\gtrsim 1$ \citep{Alvarez14}, therefore setting the value of $\sigma_{\fNLL}$ we want to achieve with probes on local PNG. A significant detection of $\fNLL$ will automatically rule out all single field models, whereas $\sigma(\fNLL)\le1$ will exclude a large number of multi-field scenarios.  
Measurements of the Cosmic Microwave Background (CMB) by the Planck satellite have put the tightest constraints on local PNG\citep{PlanckfNL}, $\fNLL = -0.8 \pm 5$.
Unfortunately we have mostly saturated the information content in the CMB, and any further improvement will come from the late time distribution of galaxies or any other tracers of the Large Scale Structure (LSS) of the Universe. The scope of this work is to present a novel way to estimate PNG using galaxy positions.
 
PNG affect the dark matter distribution at the late times in multiple ways, from the abundance of massive clusters to the clustering of galaxies $n$th-point functions, see \citep{Desjacques10,Liguori10,Alvarez14} and references therein for a review.
Next generation of galaxy surveys are expected to improve the errorbars on local PNG, DESI~\cite{DESI} and Euclid~\cite{Euclid} should get down to $\sigma_{\fNLL}\simeq 5$ using power spectrum measurements, and a combination of power spectrum and bispectrum in optical surveys could achieve $\sigma_{\fNLL}\simeq 1$ \citep{Spherex,Dyonisis18}. Recently \citep{Schmittfull18} has also shown that a combination of LSST galaxies with CMB data has similar constraining power on PNG.
For PNG constraint with intensity lines surveys using CO and CII emission lines see instead \cite{Azadeh2018}. 

Most of the aforementioned analyses rely on the unique signature of PNG in the LSS represented by the scale dependent linear bias\citep{Dalal08,Slosar08,MV2008}. 
In the presence of local PNG the relation between the galaxy and the underlying dark matter field receive a contribution on large scales absent in a Gaussian Universe
\begin{equation}
\label{eq:b1k}
\delta_g = b_g \delta_m \quad, \quad b_g = b_1 + \fNLL b_\phi \alpha(k)
\end{equation}
where the new bias parameter can be related to the logarithmic derivative of the galaxy number density with respect to $\sigma_8$, the variance of the linear power spectrum on $8 \Mpc$ scale, via \citep{Slosar08}
\begin{align}
\label{eq:bphi}
b_\phi =  \frac{\mathrm{d} \log \bar{n}}{\mathrm{d} \log \sigma_8}\;.
\end{align}
Notice that $b_\phi$ is independent of scale. We have also defined the following transfer function from the primordial potential to the density field,
\begin{equation}
\alpha(k) = \frac{3 \Omega_m H_0^2}{c^2 k^2 T(k) D(z)}
\end{equation}
with $c$ the speed of light, $H_0$ the present day Hubble constant, $T(k)$ the matter linear transfer function and $D(z)$ the linear growth factor normalized to $1/(1+z)$ in the matter dominated area.
The non-Gaussian correction is generated by the coupling between long and short scales generated during inflation, that modulates the mean number density of galaxies as a function of the long-wavelength modes. Since at low $k$ the transfer function goes to unity one expects the non-Gaussian signal on large scale to go as $k^{-2}$. 
Equation \ref{eq:b1k} has been extensively tested in numerical simulations, and overall good agreement is found with analytical calculations\citep{Desjacques08,Sefusatti2012,Biagetti17}. A further simplification is usually made in Equation \ref{eq:b1k}, that the mass function is universal, \ie$
\label{eq:b1uni}
 \mathrm{d} \log \bar{n}/\mathrm{d} \log \sigma_8 = \delta_c (b_1-1)$,
whith $\delta_c=1.686$. This is only an approximation, \cite{Biagetti17} found it to be accurate within 20\% of the measurements in the simulations, but it is useful to get a rough idea of the amplitude of signal.

Despite the new signatures in the distribution of LSS, measuring scale dependent bias in galaxy surveys is quite challenging, for two main reasons. The first one is that large scales in galaxy surveys are usually the most affected by systematic effects, for instance an incomplete knowledge of the window function or residual foreground contamination \citep{Pullen2013}. The second one is that large scales measurements intrinsically have the largest noise because of sample, or cosmic, variance.
Whereas the first kind of issues we can hope to solve for in the near future with better measurements and modeling, the latter is a much more severe problem as it is directly related to the fact we observe only one realization of the Universe.
A possible solution has been proposed by \cite{Seljak09}, who noticed that cross correlation of different tracers can reduce the effect of cosmic variance on an estimate of $\fNLL$. The idea is that two tracers in the same region of the sky will be sampling the same realization of the underlying density field, so cosmic variance can be cancelled.

For this cross correlation method to work however one would need the other main source of noise in the galaxy power spectra, \ie the shot-noise arising from the discrete number of tracers, to be negligible compared to the signal, a condition very hard to achieve in data even for a single tracer. Multitracer techniques are also complicated by the fact one has to find two galaxy populations in the same region of the sky with very different linear biases but similar very low shot-noise~\citep{dePutter2014,Ferraro2015,Gleyzes2017}
\footnote{See \cite{Schmittfull18} for a multitracer analysis using CMB lensing as the second tracer}.
Even observing the full 3D dark matter distribution and all the halos above $~3\times 10^{12} \,\Msun$ at redshift $z=1$ one barely reach $\sigma_{\fNLL} \simeq 1$, in an extremely large volume of $50 \,(\Gpc)^3$~\citep{Hamaus11,Gleyzes2017}.

The scope of this work is to propose an alternative route to sampling variance cancellation, using only a single tracer with zero or slightly negative $b_1$. The idea, as we will describe in more details in Section \ref{sec:SN} and Section \ref{sec:Forc}, is that for such a field the sampling variance itself is negligible and the only source of noise is represented by the finite number of tracers. 

For the remainder of the paper we will use Planck best fit cosmological parameters\citep{Planck} with a fiducial value of $\fNLL = 0$.

\section{Signal to noise and sampling variance cancellation}
\label{sec:SN}
Let's start with a signal to noise estimate of the amplitude of the scale dependent bias in the galaxy power spectrum using a Fisher formalism. For illustration purposes only we will assume $b_\phi$ follows the universality relation. Given a model for the covariance $C(k)$ of the signal and the noise of the power spectrum we have \citep{Tegmark}
\begin{equation}
\label{eq:Fisher}
F_{ab} = V \int \dk \frac{1}{2}\mathrm{Tr}\left[ C^{-1}\frac{\partial C}{\partial \theta_a} C^{-1}\frac{\partial C}{\partial \theta_b}\right]_{\big{|}_{\text{fid.}}}
\end{equation}
for the Fisher information of any pair of parameter $\theta_{a,b}$ in a cosmological volume $V$. The parameters covariance matrix is simply given by $\sigma_{ab} = (F^{-1})_{ab}$.
For local PNG the covariance reads
\begin{align}
C(k,z) = &P_{gg}(k,z) + \frac{1}{\bar{n}(z)} \notag \\ &= [ b + \fNLL \delta_c(b-1)\alpha(k)]^2 P(k,z) + \frac{1}{\bar{n}(z)}
\end{align}
where we have assumed the analysis is carried out in real space and from now one we will use $b$ for the fiducial value of linear bias. Let us also assume linear bias is perfectly known as well as the number density of objects, \ie $\fNLL$ is the only free parameter.
In this case the error goes as
\begin{align}
\label{eq:fNLsingle}
\sigma_{\fNLL}^{-2}=F_{\fNLL \fNLL} \propto \frac{b^2(b-1)^2 \alpha(k)^2 P^2(k,z)}{\left(b^2 P(k) + \dfrac{1}{\bar{n}}\right)^2}\;.
\end{align}
Consider now two tracers with  the same number density $\bar{n}$, and the same PNG signal, \ie the same value of $(b-1)^2$, but one with positive and one with negative $b$, \eg $b=3$, a rare cluster, and $b=-1$, a void \citep{Sheth,Hamaus2014}. 
The goal is to make the value of Equation \ref{eq:fNLsingle} as large as possible.
In the limit where shot-noise dominates over the cosmic variance we have
\begin{equation}
F_{\fNLL \fNLL} \longrightarrow \delta_c^2 b^2(b-1)^2 \alpha(k)^2 \bar{n}^2P^2(k,z)
\end{equation}
and we see that the tracer with positive bias will yield the smaller errorbar at fixed number of objects.
The other limiting case, where the shot-noise is negligible compared to sample variance, is however very different
\begin{equation}
\label{eq:bzeroF}
F_{\fNLL \fNLL} \longrightarrow \frac{\delta_c^2 (b-1)^2 \alpha(k)^2}{b^2}
\end{equation}
and the negative bias tracer performs much better than the positive one.
In a realistic analysis one always marginalizes over bias factors, which implies the Fisher matrix \emph{in real space} is singular and therefore not invertible for a fiducial $b=0$, while Equation \ref{eq:bzeroF} suggests the error on $\fNLL$ tends to zero.
\begin{figure}
\centering
\resizebox{\columnwidth}{!}{\includegraphics{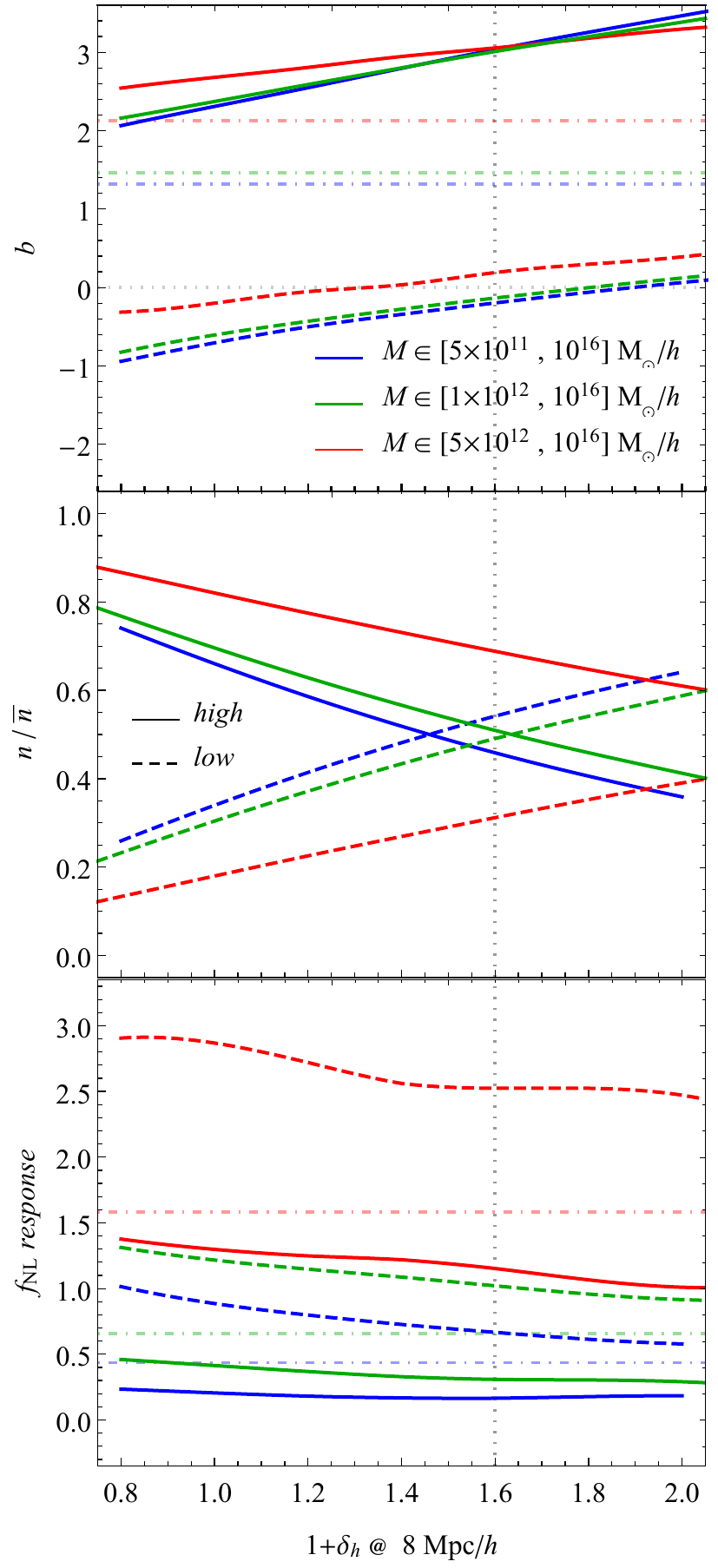}}
\caption{The large scale bias, number density and $\fNLL$ response of halos measured in the N-body simulations, as a function of the environment defined at $R_E=8\,\Mpc/h$. The upper panel shows the bias for the high, continuous lines, and low, dashed lines, samples. The three mass bins are displayed with blue, red, and green lines for $M_{min}>5\times10^{11}\,,1\times10^{12},\,5\times10^{12}\Msun/h$ respectively. The bias of the parent sample is shown with dot-dashed lines. The mid-panel shows the number density of the high and low samples in units of parent sample one. The bottom panel displays the $\fNLL$ response measured using Equation \ref{eq:bphi} in the high and low fields compared to the response of the full catalog (dot-dashed lines)}\label{fig:bias}
\end{figure}
It is also important to keep in mind that for $b>0$ the cosmic variance limit is reached at very low number densities, $\bar{n}\simeq 10^{-4}$, after which there is no more improvement for a single tracer\citep{Hamaus11,Ferraro2015,Gleyzes2017}.

\begin{figure}
\centering
\resizebox{\columnwidth}{!}{\includegraphics{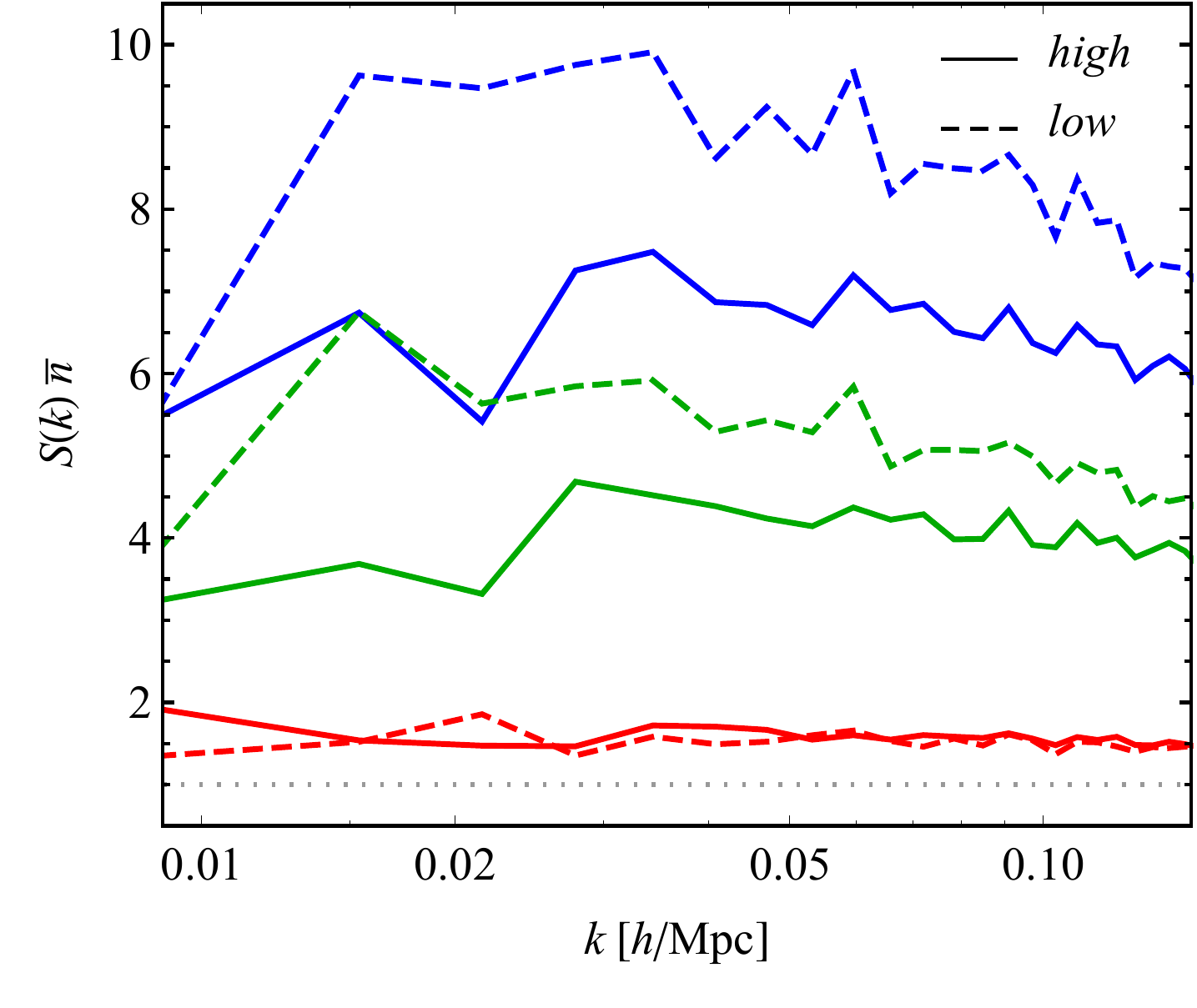}}
\caption{Measurements of stochasticity in the high and low fields multiplied by the corresponding number density $\bar{n}$. Colour coding as in \fig{fig:bias}.
The horizontal dotted black line shows the Poisson expectation.}\label{fig:noise}
\end{figure}

\section{A zero bias field}
\label{sec:sims}
The analysis of the previous section suggests that, in the low shot noise limit and at fixed $\fNLL$ response, the closer to zero the bias is the better we can constraint local PNG. 
This is saying that another way to cancel sample variance would be to have zero power on large scales, such that the only signal left is in PNG.
This is a very special feature of scale dependent bias, as the broadband power does not carry any information about $\fNLL$ and the fiducial value is $f_\mathrm{NL}=0$.
But how do we get a tracer with $b=0$?
Galaxy bias for halo mass selected samples is never below 0.6 \cite{SeljakWarren04} and galaxy 
samples of the current and next generation of redshift surveys will all have bias larger than one and therefore are not an option. A LSS tracer however can be defined with more than just one number, \eg mass for halos or observed flux/luminosity for galaxies, and we can use other criteria to construct our sample. 
The classic example of negative bias tracers are voids \citep{Sheth}, commonly identified as underdense region in a galaxy distribution. Unfortunately voids often have large negative bias \citep{Hamaus2014}, $b<-1$, and extremely low number densities, hence they are not very useful to constraint PNG.

A simple selection can be done using local density: if galaxies live in a dense environment they will be more biased than the galaxies in a low density environment, even for the same luminosity \cite{Abbas2007,Pujol,Paranjape2017,Shi2017,Salcedo2018,Paranjape2018,Alam2018,Han2018}. One can show that 
besides halo mass the environment is the main contributor that sets the bias \citep{Shi2017,Castorina18}.
Since in a real survey one does not have access to the 3D dark matter field we will define the local density as the value of the mass weighted \textit{halo} field in a sphere of radius $R_E$. The choice of the environmental scale is driven by two competing effects.  First, $R_E$ should be large enough to avoid noise coming from the sparsity of the sample to affect our estimate of the environment. At the same time we want also to minimize $R_E$.
Our selection is still local, as long as $R_{E}$ is 
not too large. This means that on scales larger than $R_{E}$ the bias is scale independent, while on scales comparable to $R_{E}$ it becomes scale dependent because 
the one and two halo terms becomes sensitive to the halo profile, where the "halo" is of size $R_E$.
A smaller value of $R_E$ will also ensure the noise is scale independent on large scales. 

To test this idea we have run a set of N-body simulations of a Planck cosmology using the GADGET code \citep{Springel}, with box size $L=500\,\Mpc$ and $1024^3$ particles.
We then found Friends-of-Friends halos and divided the full halo catalogs in three sample with different minimum mass $M_{min}=5\times 10^{11},\,1\times 10^{12},\,5\times 10^{12}\,\Msun/h$. The corresponding comoving number densities are $\bar{n} = 6.4,\,32,\,78\,\times 10^{-4} \,[h/\text{Mpc}]^3$. 
Each sample is then further divided according to the following environmental criterion: for a fixed threshold value of the parent halo density field at $R_E$, halos who live in regions above the threshold form one sub-sample, called \textit{high} bias halos, and all the others make the \textit{low} bias sub-sample. We have repeated the measurements at $R_E = 6,\,8,\,10\,\Mpc$, finding similar conclusions. 

Results are shown in \fig{fig:bias} for $R_E=8\,\Mpc$ and $z=1$.
The upper panel present the measurement of the bias as a function of the environment. The bias values have been obtained from a fit at low $k$ of $P_{hm}(k)/P_{mm}(k)$, where $h=\{\text{low,high}\}$, as a function of the environmental threshold.  Errorbars are too small and will not be displayed. For reference the original samples have bias $b=1.32,\,1.48,\,2.13$, shown as dot-dashed lines (blue, green and red respectively).
We notice that once split by the environment, all the halos have very similar bias irrespective of their mass\citep{Pujol,Shi2017,Castorina18}.
When the density is around $1.6$ we can identify a sample with zero bias, and the high field has bias of roughly three.

In a Fisher analysis it is also important to quantify the noise in the high and low samples. The middle panel shows the fraction of the initial halos that end up in the two subsamples. For $M_{min}>5\times 10^{12}\,\Msun/h$, only a 30\% of the original sample has zero bias, while denser initial samples, \ie lower $M_{min}$, can host 45-55\% of unclustered halos.
However, despite the fact the noise in the halo or galaxy field is usually consider Poissonian (but see \citep{Hamaus2010,Baldauf13,Hand2018} for why this is often a bad assumption), the high and low bias fields will certainly deviate from the Poisson regime as a result of the large exclusion region imposed at $R_E$. 
To better measure the noise we have therefore computed stochasticity between the high/low bias fields and the dark matter
\begin{align}
S_h(k) = P_{hh}(k)-\frac{P_{hm}^2(k)}{P_{mm}(k)}\,.
\end{align}
In the low-$k$ limit the above expression should approach $1/\bar{n}$ if halos are a Poisson process. As we want to probe the large scale limit of the noise we ran additional simulations of the fiducial cosmology in a box of $L=1000\,\Mpc$ with $N=1680^3$ particles.
We do indeed find in \fig{fig:noise} large deviations from the simple shot-noise, in both the high and the low sub-samples. This will affect our Fisher calculation, thus we fit for $S(k)$ at low $k$ and consider the constant term as a renormalized shot-noise value.

Using the $\Lambda$CDM simulations we can also measure the response to $f_\mathrm{NL}$, using Equation \eqref{eq:bphi}. This is indeed the very definition of scale dependent bias and it is more accurate than measuring it from simulations that include PNG.
For this purpose we ran an additional set of simulation with slightly different value of the fiducial value of $\sigma_8^{fid}$, we picked $\sigma_8^{\pm}=0.833\pm 0.02$, and then took the numerical derivative according to Equation \eqref{eq:bphi}. The results are shown in the bottom panel of \fig{fig:bias}. Compared to the horizontal lines, which correspond to the response of the three parent samples, we find that zero bias tracers have in general larger response. Conversely the high bias field is less sensitive to PNG. Since environment and formation time are strongly correlated \citep{Han2018}, \ie recently formed halos live in denser environments, our results are qualitatively in agreement with \citep{Slosar08,Readi2010}, who found that old (young) halos have smaller (larger) $\fNLL$ response than average. 
An analytical investigation of these results using excursion sets peaks \citep{ESP} is work in progress.
\section{Forecast}
\label{sec:Forc}

\begin{figure}
\includegraphics[width=1\columnwidth]{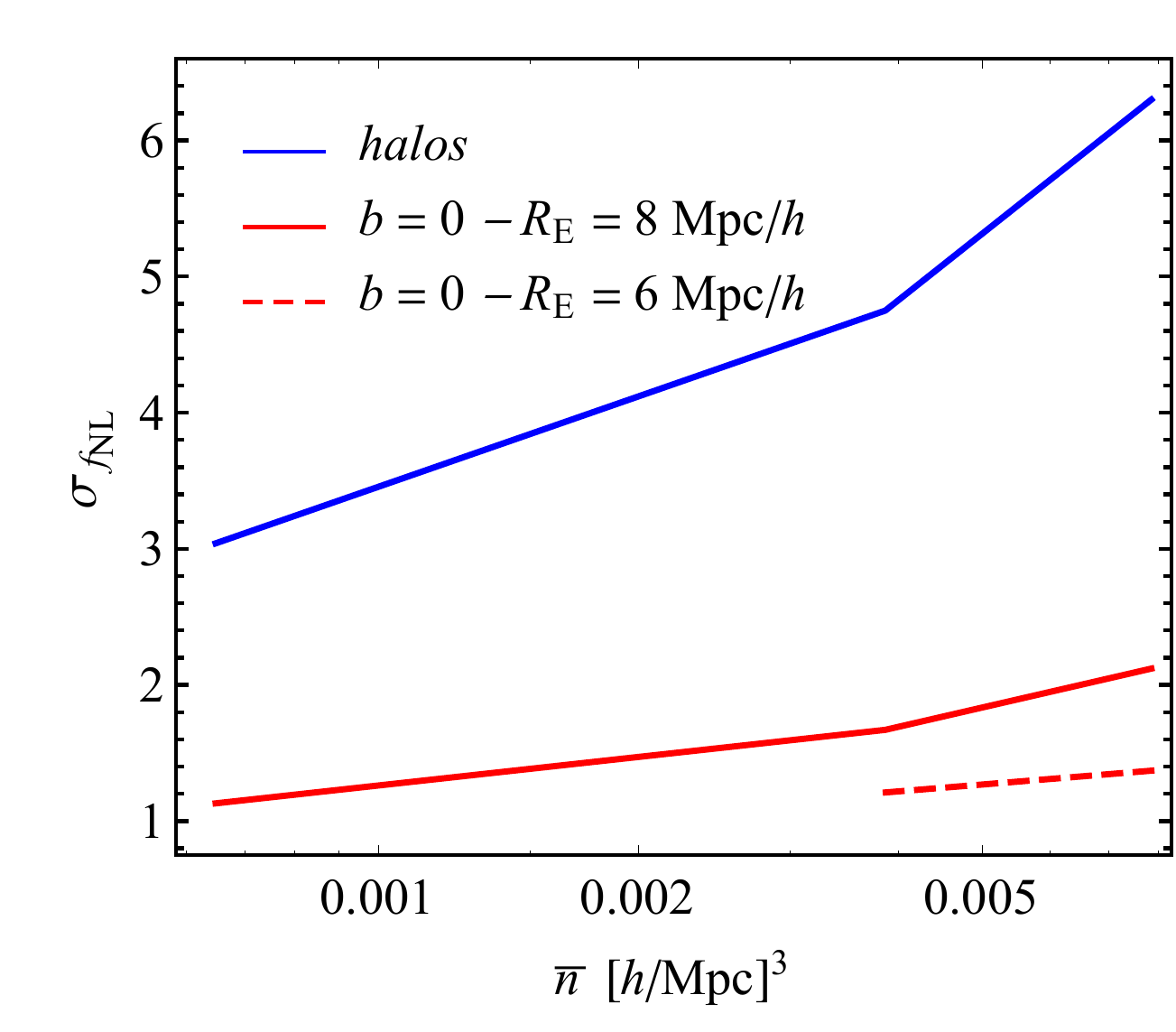}
\caption{Error on $\sigma_{\fNLL}$ for a survey of volume $V=50\,(\Gpc)^3$ at $z=1$. The standard analysis for a single halo population is in blue, while our new method, combing the high and low bias fields, is shown with the red continuous line. The dashed red line shows the case of $R_E = 6\Mpc/h$.}\label{fig:sigma}
\end{figure}
In the previous section we presented measurements in the simulations of the signal and the noise required to perform a Fisher analysis of PNG. 
We consider an hypothetical case of a survey at $z=1$ in a $V=50\,(\Gpc)^3$ volume and forecast the error on $\sigma_{\fNLL}$ for the three mass bins discussed in the previous section. Our Fisher matrix has two free parameters, the linear bias $b$ and $\fNLL$, and we include all the modes from $k_{min} = 2\pi /V^{1/3}$ and $k_{max} =0.075\,\kMpc$.
\fig{fig:sigma} is the main result of this paper. 
The blue line is the standard single tracer analysis, for which we find higher number densities yield worse errorbars. This happens because for halos, \ie positive biased tracers, we are in the sample variance dominated regime, and high bias wins over high number densities. 

For the zero bias fields we assume measurements of the auto power spectrum of the two samples, as well as their cross-correlation.
The improvement over the parent samples is dramatic, the gain is a factor of three or larger for the mass bins considered in this paper.
Whereas one would expect that high number densities will always be better for zero bias tracers, our analysis suggests this is not the case. The reason is that the noise in the low bias sample is much larger than the Poisson value, see \fig{fig:noise}, making the zero bias field less constraining for dense samples. The red dashed lines presents the same analysis for $R_E=6\,\Mpc/h$. The sample with $M_{min}=5\times 10^{12} \,\Msun/h$ is too sparse to return a sensible measurement of the environment, and it is therefore not shown. The constraint improves over the case of $R_E =8\,\Mpc/h$, mostly due to the lower noise levels, according to the discussion in Sec. \ref{sec:sims}.
It would be interesting to see if other ways of selecting zero bias tracers yield different constraints.
For $\bar{n}\simeq 10^{-3}\,(\text{Mpc}/h)^{-3}$ at $z=1$ we find $\sigma_{\fNLL}\simeq1$. 
Compared to the analysis in \citep{Hamaus11,Gleyzes2017}, to achieve the $\sigma_{\fNLL}\simeq1$ we did not have to assume we have information from the dark matter density field.

\section{Conclusions}
In this paper we have shown how zero bias tracers of the LSS could improve our knowledge of local PNG. Our analysis takes advantage of the fact that for such tracers cosmic variance can be made arbitrarily small and the only source of variance in a measurement of the power spectrum becomes the shot-noise. 
Given a halo catalog, we proposed and tested in simulations a simple method to select such a tracer, using environmental information obtained from the original sample itself. We took particular care in defining the noise of this new field, showing that it significantly deviate from the Poisson value.
Using measurements in the simulations we then forecasted the error on $\fNLL$ and found factor of three improvement in $\sigma_{\fNLL}$ over standard positive bias tracers, for a variety of halo mass thresholds.
Further gains in constraining power could be obtained by optimal weighing the halos or galaxies according to \citep{Hamaus11}.
A real data analysis will be done in redshift space, where one could exploit the fact that the power spectrum becomes proportional to $b+f(z)\mu^2$, where $f(z)$ is the linear growth factor and $\mu$ is the cosine of angle between the the mode $\mathbf{k}$ and the line of sight to the galaxies. Indeed for some value of $\mu$ we could have
$b\simeq-f\mu^2$.
In this respect the two tracers one needs for the sampling variance cancellation technique of \cite{Seljak09} were already there in the first place, as one can use the real density field and the velocity field generating RSD.
One could also imagine, with enough galaxies at hand, to optimize the analysis to cancel cosmic variance in several $\mu$-wedges.
We should however keep in mind that in redshift space non-linear selections of galaxies, like the one discussed in this paper, generate velocity bias \citep{Seljak2012,Chuang2017}. We plan to return to the case of redshift space distortions in a forthcoming paper.

\textit{Acknowledgments.} E.C. thanks Simone Ferraro, Marko Simonovic and Dan Green for useful discussions and  Joanne Cohn for comments on a earlier version of the draft. We acknowledge support of NASA grant NNX15AL17G. 
\bibliographystyle{h-physrev}
\bibliography{references}

\end{document}